\documentclass[aps,pra,reprint,showpacs,nofootinbib,superscriptaddress]{revtex4-1}
\input{Qcircuit}
\usepackage{graphicx,comment}
\usepackage{amsfonts}
\usepackage{amsmath,amssymb}
\usepackage{bm}
\usepackage{color}
\usepackage{epstopdf}
\usepackage{epsfig}
\usepackage[font=small,
   justification=justified,
   format=plain]{caption} 
\usepackage{subfig}
\usepackage{float}
\usepackage{verbatim}
\usepackage{hyperref}
\usepackage{multirow}
\usepackage[table,xcdraw]{xcolor}
\usepackage{tabularx}
\usepackage{wrapfig}
\hypersetup{
     colorlinks   = true,
     citecolor    = blue
}
\usepackage{lineno}
\usepackage[thicklines]{cancel}
\usepackage{color, colortbl}
\definecolor{LightCyan}{rgb}{0.88,1,1}
\usepackage{url}
\usepackage{dcolumn}
    
{\rm }

\def\be{\begin{equation}}
 \def\ee{\end{equation}}
 \def\bea{\begin{eqnarray}}
 \def\eea{\end{eqnarray}}

\def\2{\frac{1}{2}}
\def\4{\frac{1}{4}}

\begin{document}
\title{Collective Neutrino Oscillations on a Quantum Computer}

\author{K\"ubra Yeter-Aydeniz }
\email{yeteraydenik@ornl.gov}
\affiliation{Physics Division, Oak Ridge National Laboratory,
  Oak Ridge, TN 37831, USA}

\author{Shikha Bangar}
\email{sbangar@vols.utk.edu}
\affiliation{Department of Physics and Astronomy,  The University of Tennessee, Knoxville, TN 37996-1200, USA}

\author{George Siopsis}
\email{siopsis@tennessee.edu}
\affiliation{Department of Physics and Astronomy,  The University of Tennessee, Knoxville, TN 37996-1200, USA}

\author{Raphael C.\ Pooser}
\email{pooserrc@ornl.gov}
\affiliation{Computational Sciences and Engineering Division, Oak Ridge National Laboratory,
  Oak Ridge, TN 37831, USA}

\date{\today}

\begin{abstract}

We calculate the energy levels of a system of neutrinos undergoing collective oscillations as functions of an effective coupling strength and radial distance from the neutrino source using the quantum Lanczos (QLanczos) algorithm implemented on IBM Q quantum computer hardware. Our calculations are based on the many-body neutrino interaction Hamiltonian introduced in Ref.\ \cite{Patwardhan2019}. 
We show that the system Hamiltonian can be separated into smaller blocks, which can be represented using fewer qubits than those needed to represent the entire system as one unit, thus reducing the noise in the implementation on quantum hardware. 
We also calculate transition probabilities of collective neutrino oscillations using a Trotterization method which is simplified before subsequent implementation on hardware. These calculations demonstrate that energy eigenvalues of a collective neutrino system and collective neutrino oscillations can both be computed on quantum hardware with certain simplification to within good agreement with exact results.
{\footnote{This manuscript has been authored by UT-Battelle, LLC, under Contract No. DE-AC0500OR22725 with the U.S. Department of Energy. The United States Government retains and the publisher, by accepting the article for publication, acknowledges that the United States Government retains a non-exclusive, paid-up, irrevocable, world-wide license to publish or reproduce the published form of this manuscript, or allow others to do so, for the United States Government purposes. The Department of Energy will provide public access to these results of federally sponsored research in accordance with the DOE Public Access Plan.}}

\end{abstract}

\maketitle

\section{Introduction}
According to the Standard Model of particle physics, neutrinos are neutral  elementary particles that exist in three leptonic flavors: electron, muon, and tau neutrino. Recent experimental evidence \cite{Nobel, Super, SNO, Daya} indicates that neutrinos have multiple mass eigenstates which do not coincide with flavor eigenstates. This  causes mixing between different neutrino flavors. Thus, neutrinos have been observed to oscillate between different flavors as they propagate, which is a quantum mechanical phenomenon.

Early research in neutrino oscillations boosted observations of the flux of solar neutrinos \cite{Solar}, experimental studies of neutrino beams, as well as  theoretical research in the nature of neutrinos that appear to be emitted and absorbed in flavor eigenstates but travel as mass eigenstates. 
After studying neutrino flavor transitions, it has been realized that neutrinos experience self-maintained coherent oscillations \cite{CollectiveNO}. Moreover, these collective neutrino oscillations are not disrupted by a homogeneous and isotropic environment. This phenomenon was studied in detail in connection with early universe scenarios \cite{Abazajian2002} and supernovae \cite{Wu2014}.  

In the astrophysical events mentioned above, an abundant number neutrinos are created. For example, in a core-collapse supernova event the energy released is of order $\sim 10^{59}$ MeV and $99\%$ of this energy is carried away by neutrinos and antineutrinos. This large amount of of energy corresponds to $\sim 10^{58}$ neutrinos~\cite{Martinez-Pinedo2016}.
Studying interactions between this many neutrinos using classical computers is a daunting task. One needs to account for strong correlations due to collective neutrino oscillations. Calculating the energy spectrum of the collective neutrino system is central to understanding the physics behind neutrino interactions giving rise to a collective behavior. In Ref.~\cite{Patwardhan2019}, a mean-field approximation was used to calculate the eigenvalues and eigenstates of a collective neutrino system.  They introduced a method to systematically obtain the eigenvalues and eigenstates of a many-body Hamiltonian describing collective neutrino oscillations. Remarkably, using this method, one can perform calculations in a 15-neutrino system with a personal computer. 


On the other hand, the exponential speedup that fault-tolerant quantum computers provides hope to simulate and understand high energy physics that is inaccessible to classical computers. Recently, there has been exciting progress, for example, towards understanding parton showers using quantum computers~\cite{Nachman2021}. A similar speedup with quantum computers is also expected in the study of collective neutrino oscillations. The first attempt to study neutrino oscillations on a quantum processor was discussed in~\cite{Noh2012} where a scheme to study the dynamics of neutrino oscillations on a trapped-ion quantum computer was introduced. The first experimental results on the dynamics of neutrino oscillations employing quantum hardware were presented in Ref.\ \cite{Arguelles2019}. Two- and three-flavor neutrino oscillations were represented on a superconducting quantum processor, and the survival probability of different neutrino flavors was studied as a function of neutrino energy. Here, we make use of the quantum Lanczos and hubrid quantum-classical algorithms to obtain energy spectra for the Hamiltonian in the mass basis of neutrino systems undergoing oscillations, in addition to transition probabilities. 

Concurrently with the present work, Ref.~\cite{Hall2021} also studied the simulation of collective neutrino oscillations on a quantum computer. The authors of that study explore the real-time many-body evolution of a collective neutrino system using first-order Trotterization combined with error mitigation, a complementary technique to the Quantum Lanczos (QLanczos) method used in the present work. 


Our experimental results for the energy eigenvalues of $N=4$ collective neutrino system are in very good agreement with exact diagonalization, 
enabling us to observe flavor oscillations in the collective neutrino system with $N=3$ and $N=4$ neutrinos.

The layout of this paper is as follows. In Sec.\ \ref{sec:model} we introduce the collective neutrino system to be implemented on a quantum computer. Next, in Sec.\ \ref{sec:qc} we discuss how the neutrino system can be studied with a quantum computer, including the methods for the calculation of the energy levels of the system and the transition probabilities between neutrinos of different flavors. In Sec.\ \ref{sec:res} we present our results for the energy levels of a four-neutrino system as functions of the coupling parameter that depends on the radial distance from the neutrino source using the QLanczos algorithm. We also discuss the calculation of transition probability amplitudes for 3-, and 4-qubit collective neutrino systems as a function of time using a simplified first-order Trotterization method. Finally, in Sec.~\ref{sec:con} we summarize our conclusions and outlook.

\section{The Model}
\label{sec:model}
Neutrino interactions can be described by a many-body Hamiltonian which consists of neutrino oscillations in vacuum, interactions of neutrinos with background matter and other neutrinos and anti-neutrinos. In Ref.~\cite{Patwardhan2019}, eigenvalues and eigenstates of the many-body collective neutrino oscillation Hamiltonian were studied after applying certain simplifications such as limiting to two flavor/mass states of neutrinos, ignoring interactions with anti-neutrinos as well as the background matter. With these simplifications, the Hamiltonian representing the vacuum and neutrino self-interaction terms can be written as
\be
H = \sum_{{\bm{p}}} \omega_{{\bm{p}}} \Vec{B}\cdot \Vec{J}_{{\bm{p}}}+\frac{\sqrt{2}G_F}{V}\sum_{{\bm{p,q}}}(1-\cos \vartheta_{{\bm{pq}}})\Vec{J}_{{\bm{p}}}\cdot \Vec{J}_{{\bm{q}}}~,
\ee
where $\Vec{B}=(0,0,-1)_{\text{mass}}=(\sin{2\theta},0,-\cos{2\theta})_{\text{flavor}}$, $\theta$ is the vacuum mixing angle, and $\omega_{{\bm{p}}}=\delta m^2/(2|{\bm{p}}|)$ are the vacuum oscillation frequencies. The neutrino mass-basis isospin operators $\Vec{J}_{{\bm{p}}}$ can be written in terms of fermionic creation and annihilation operators as
\be
J_{{\bm{p}}}^+=a_1^\dagger({\bm{p}})a_2({\bm{p}})~,
\ee
\be
J_{{\bm{p}}}^-=a_2^\dagger({\bm{p}})a_1({\bm{p}})~,
\ee
\be
J_{{\bm{m}}}^z=\frac{1}{2}\left(a_1^\dagger({\bm{p}})a_1({\bm{p}})-a_2^\dagger({\bm{p}})a_2({\bm{p}})\right)~.
\ee
In the case where neutrinos are assumed to be emitted isotropically from a single spherical emission surface, the neutrino self-interaction term can be simplified to
\be
H \approx \sum_{{\bm{p}}} \omega_{{\bm{p}}} \Vec{B}\cdot \Vec{J}_{{\bm{p}}}+\mu(r) \left( \sum_{{\bm{p}}}\Vec{J}_{{\bm{p}}} \right)^2 \label{eq:Ham}
\ee
where $\mu(r) =\frac{G_F}{\sqrt{2}V} \left(1-\sqrt{1-\frac{R_v^2}{r^2}}\right)^2$,
with $G_F$ the Fermi interaction constant, $V$ the quantization volume, $r$ the distance from the center of a neutrino sphere of radius $R_\nu$, and $\bm{p}$ the index labeling the oscillating frequencies present in the system. 

\section{Quantum Computation}
\label{sec:qc}
We use the hybrid quantum-classical QLanczos algorithm discussed in \cite{Motta2019, YeterAydeniz2020} to find the eigenvalues of the collective neutrino many-body interaction Hamiltonian \eqref{eq:Ham}. To study a Hamiltonian on a quantum computer one needs to express it in terms of Pauli spin matrices. Since neutrinos are fermions, observables can be represented using a two-dimensional $SU(2)$ algebra. The isospin operators can be expressed in term of Pauli matrices as $J_i= \frac{\sigma_i}{2}$, where $\sigma_i \in \{X, Y, Z \}$ are Pauli spin matrices. Therefore, the neutrino interaction Hamiltonian \eqref{eq:Ham} in mass basis can be written as
\be
H=-\frac{1}{2}\sum_{p=1}^M\omega_p Z_p+\frac{\mu(r)}{4} ( \bm{X}^2 + \bm{Y}^2 + \bm{Z}^2 )~,\label{eq:PauliHam}
\ee
where $p$ labels the oscillation frequencies present in the system, $\bm{X} = (X_1,\dots ,X_M)$, and similarly for $\bm{Y}$ and $\bm{Z}$. 


We choose a system with $N$ neutrinos distributed evenly across $M$ oscillation frequencies, and for simplicity we set $M=N$. In this system, one needs to calculate $2^N$ eigenvalues, which becomes an intractable problem on classical computers as the number of neutrinos grows. Quantum computers provide an exponential speedup in the calculation of eigenvalues and eigenvectors as functions of $\mu/\omega_0$, where the oscillation frequencies are given by $\omega_p=p\omega_0$ \cite{Cervia2019}. We use the numerical parameters for dense neutrino gasses provided in Ref.\ \cite{Cervia2019}.

The system Hamiltonian \eqref{eq:PauliHam} in the mass basis is separable into independent blocks. This follows from the fact that $H$ commutes with the total number operator
\be \mathcal{N} = \sum_{p=1}^N \frac{I - Z_p}{2}. \ee
Since $\mathcal{N}$ has $N+1$ distinct eigenvalues, $0,1,\dots, N+1$, the Hamiltonian splits into $N+1$ blocks that can be studied independently. The $k$th block has dimension $d_k = {N \choose k}$, and $d_0+d_1+ \dots + d_N = 2^N$. The largest block grows asymptotically as $\frac{2^N}{\sqrt{N}}$, requiring $N - \frac{1}{2} \log_2 N$ qubits. This is not a significant improvement for large $N$, but for values of $N$ relevant to NISQ devices, it provides a significant reduction in required resources.

To compute the spectra, we express each reduced block Hamiltonian as a sum of unitary matrices\ \cite{YeterAydeniz2021Chem}:
\be
H_{\text{block}}=\sum_I c_I \sigma_I
\ee
where $I= \{ i_0 , \dots, i_{n_q -1} \}$ and $\sigma_I = \sigma_{i_0}\sigma_{i_1}\dots\sigma_{n_{q}-1}$, with $\sigma \in \{I,X,Y,Z\}$ and $n_q$ the number of qubits used for this block Hamiltonian. The coefficients $c_{I}$ are found from
\be
c_{I}= \frac{1}{2^{n_{q}}}\text{Tr} [H_{\text{block}} \sigma_I]~.
\ee
Here we study the four-neutrino system Hamiltonian. The matrix elements corresponding to no particle, i.e., in the state $|0000\rangle$, and four particles, i.e., in the state $|1111\rangle$, form  one-dimensional blocks, so the eigenvalues are readily available analytically. The subspaces of single-particle states, $\{ |1000\rangle, |0100\rangle, |0010\rangle, |0001\rangle \}$ and three-particle states, $\{ |1110\rangle, |0111\rangle, |1011\rangle, |1101\rangle \}$ form independent four-dimensional blocks each of which can be represented by a two-qubit system. 

The largest block corresponds to the six-dimensional subspace of two-particle states, $\{ |1100\rangle, |0101\rangle, |1010\rangle, |1001\rangle, |0011\rangle, |0110\rangle \}$ that can be studied using a three-qubit system. 
They form a block Hamiltonian that can be written as
\begin{eqnarray}
H_{\text{block}} &=& \frac{(\omega_0 + 6 \mu)}{4} I+\frac{\mu}{2} H_1 +H_2 \label{3qubitreducedHam}
\end{eqnarray}
with 
\begin{equation}
    \begin{split}
        H_1= &X_0 X_1 X_2 +  Z_0 X_1 X_2 + Y_0 Y_1 X_2 + X_0 X_1 +  X_0 Z_1 \\& + Z_0 X_1 +Y_0 Y_1 +X_1 X_2 + Z_1 X_2 -Z_0Z_1 \\& + X_0 + X_1 + X_2  ~,
    \end{split}
\end{equation}
\begin{equation}
\begin{split}
        H_2= &\frac{(2\omega_0-\mu)}{2}Z_0  Z_1  + \frac{\omega_0}{4} Z_1 Z_2  + \frac{\mu+\omega_0 }{2}Z_0\\& +  \frac{2\mu-\omega_0 }{4} Z_1 + \frac{\omega_0}{4} Z_2~,
        \end{split}
\end{equation}
where we added arbitrary values (additional energy levels $E_1', E_2'$) as the last two elements of the diagonal to create a matrix for a 3-qubit system. One needs to make sure that these additional energy levels are not at the end of the energy spectrum, or choose an initial state which is orthogonal to the corresponding eigenvectors, otherwise the QITE algorithm may converge to these spurious states. We set $E_1' = E_2' =\omega_0$. 

Similarly,
\be
\begin{split}
H_{\text{block}}^{\pm}=& \frac{6\mu\pm 5 \omega_0}{2} I +\mu(X_0+X_1+X_0 X_1)\\&+\omega_0 (Z_0+\frac{1}{2}Z_1)~, 
\end{split}
\ee
 gives the reduced Hamiltonian blocks for 1-particle ($H_{\text{block}}^{-}$) and 3-particle ($H_{\text{block}}^{+}$) states, respectively. 
 
 Experimentally, we have $\omega_0=1.055 \times 10^{-16}$ MeV~\cite{Cervia2019}. Here, we work in units in which $\omega_0 = 1$, thus effectively we calculate energy levels, $E/\omega_0$, as functions of $\mu(r)/\omega_0$.

\subsection{QLanczos Algorithm}
\label{sec:QL}
In this section we provide a short discussion on how we apply our hybrid quantum-classical QITE / QLanczos algorithm to calculate the energy eigenvalues of a collective neutrino system.

The QLanczos algorithm is based on the quantum imaginary-time evolution (QITE) algorithm (explained in Appendix~\ref{sec:QITE}) whose main goal is to simulate the non-unitary imaginary-time evolution of a system on a quantum computer with only unitary operations. It was proposed in Ref.~\cite{Motta2019} and, unlike other approaches~\cite{McArdle2019}, does not require ancilla qubits or classical optimization. However, it is implemented with multiple single- and two-qubit gates at each step of the imaginary-time evolution until the system converges to the ground state energy. To make the implementation more compatible with NISQ devices, we proposed certain adaptations that reduced the depth of the quantum circuits \cite{YeterAydeniz2019, YeterAydeniz2020, YeterAydeniz2021Chem}.


At each step of the QITE algorithm we calculate energy expectation values which are then used as input by the QLanczos algorithm. As menthioned earlier, to increase the performance of our algorithm, we divide the Hamiltonian for the collective neutrino system in the mass basis into independent blocks. We apply the QITE and QLanczos algorithm on each block Hamiltonian choosing initial states informed by the symmetry of the block in order to reduce the number of imaginary-time steps required for convergence.

Having found the ground state, one can in principle use the QITE algorithm to obtain excited states by selecting an initial state $|\Psi_0\rangle$ orthogonal to the ground state. However, this in practice is an impossible task on NISQ devices due to errors. Instead, we apply the quantum Lanczos (QLanczos) algorithm which uses the QITE states and energy expectation values at select steps as inputs. These states span the Krylov space $\mathcal{K}$ of the QLanczos algorithm such that ($\mathcal{K}$) spanned by $\{ |\Phi_0\rangle, |\Phi_2\rangle, \dots \}$ where $|\Phi_l\rangle=c_l e^{-l\Delta \tau H_\text{block}}|\Psi_s\rangle$ and $|\Psi_s\rangle$ is the state at the $s$-th QITE step. After filling out the Krylov space we build the overlap ($\mathcal{T}$) and Hamiltonian ($\mathcal{H}$) matrices from the energy expectation value measurements on quantum hardware. The elements of these matrices are defined as \be
\mathcal{T}_{l,l'}=\langle \Phi_l|\Phi_{l'}\rangle~, \ \ \ \ \mathcal{H}_{l,l'}=\langle \Phi_l|H_\text{block}|\Phi_{l'}\rangle~,
\ee
respectively. We need to express these quantities in terms of energy expectation values which are obtained from measurements on a quantum computer. We obtain
\be
\mathcal{T}_{l,l'}=\langle \Phi_l|\Phi_{l'}\rangle=\frac{c_l c_{l'}}{c_r^2}~,
\ee
and 
\be
\mathcal{H}_{l,l'}=\langle \Phi_l|H_\text{block}|\Phi_{l'}\rangle=\mathcal{T}_{l,l'}\langle \Phi_r|H|\Phi_{r}\rangle~,
\ee
The normalization constants are found recursively through
\be
\frac{1}{c_{r+1}^2}=\frac{\langle \Phi_r | e^{-2 \Delta \tau H_\text{block}}|\Phi_r\rangle}{c_r^2} \label{cr}
\ee
where $r=\frac{l+l'}{2}$ and $l, l'$ are even integers. For the experimental realization of these coefficients, we expand the numerator to second order in $\Delta \tau$ as $\langle \Phi_r | e^{-2 \Delta \tau H_\text{block}}|\Phi_r\rangle = 1-2 \Delta \tau \langle \Phi_r | H_\text{block}|\Phi_r\rangle+2(\Delta \tau)^2 \langle \Phi_r | H_\text{block}^2|\Phi_r\rangle+\mathcal{O}( (\Delta \tau)^3)$. This second-order expansion requires measuring energy expectation values as well as second moments, $\langle H_\text{block}^2\rangle$, at each QITE step on quantum hardware. Although this doubles the number of measurements required on quantum hardware with the method described earlier, after obtaining the QITE states from noisy simulator we can do circuit bundling and combine up to 900 (depending on the provider and backend) circuits in a single job and run these circuits on quantum hardware. For this second-order expansion, to improve the result with higher $\mu(r)$ values one needs to choose a smaller value of $\Delta \tau$ since the order of magnitude of the neutrino system Hamiltonian is directly proportional to the $\mu(r)$ value.

The next step is to solve the generalized eigenvalue equation 
\be
{\bm{\mathcal{H}x}}=E {\bm{\mathcal{T}x}}~,
\ee \label{eq:geneig}
which provides an approximation, 
\be
|\Psi[E]\rangle=c_E\left(x_0^{(E)}|\Phi_0\rangle+x_1^{(E)}|\Phi_2\rangle+\dots \right)~,
\label{eq:eigvec} \ee
to the eigenvectors of the Hamiltonian. Here, $c_E^{-1} = \| \sum_{l=0,1,\dots} x_l^{(E)} |\Phi_l\rangle \|$. The eigenvalues $E$ provide an approximation to the energy levels of the Hamiltonian. Unfortunately, these eigenvalues are numerically unstable due to the noise from quantum hardware. As the number of qubits increases in the system, the measurement outcomes become more noisy resulting in incorrect energy levels. To decide if an energy $E$ obtained from \eqref{eq:geneig} using experimental results is close to the exact energy eigenvalue of the Hamiltonian~\eqref{eq:PauliHam}, we use a minimum uncertainty criterion: we calculate the uncertainty in energy from $\Delta E=||H_\text{block} |\Psi[E]\rangle-E |\Psi[E]\rangle||$. 
We scan the eigenvalues obtained from Krylov space at each QITE step and discard the eigenvalues with uncertainty that is greater than a certain value $\delta$, i.e., we require $\Delta E \leq \delta$. We summarized the QLanczos algorithm in Fig.\ 10 of Ref.\ \cite{YeterAydeniz2020}.


We use the information from symmetry considerations of the reduced block Hamiltonian to choose different initial states, $|\Psi_0\rangle$, for each block in order to obtain all energy levels of the Hamiltonian \eqref{eq:PauliHam}.



Due to the noise in the hardware, no information is added when we add more than two QLanczos vectors to the Krylov space. Therefore, we worked with two-dimensional Krylov spaces. Using the reduced block Hamiltonian in two- and three-qubit systems, we first ran the QITE algorithm to obtain the two lowest energy eigenvalues of the symmetry sector corresponding to the initial state choice. To obtain the two highest energy eigenvalues, we ran the QITE algorithm using $-H_{\text{block}}$ with a different initial state. This yielded all energy eigenvalues of a two-qubit reduced Hamiltonian block. For the two remaining energy levels of a three-qubit Hamiltonian block (recall that two eigenvalues were added arbitrarily to supplement the six-dimensional block), we obtained the eigenstates exactly by exploiting the symmetry of the reduced Hamiltonian block and then constructed quantum circuits generating them which we implemented on quantum hardware to obtain the experimental energy eigenvalues. As it is easily seen from the three-qubit reduced block Hamiltonian in \eqref{3qubitreducedHam}, the eigenvalue $2\mu$ is degenerate and the corresponding eigenvectors are $\frac{1}{\sqrt{2}} (|010\rangle-|011\rangle)$ and $\mathcal{N} [(1+ \frac{2}{\mu} )|010\rangle - |000\rangle -2|001\rangle+|101\rangle+2|100\rangle]$, where $\mathcal{N}^{-2} = 11+ \frac{4}{\mu} + \frac{4}{\mu^2}$.

In the case of larger systems described by a larger number of qubits, in order to reach the middle energy levels, one can modify the Hamiltonian to $H_{\text{block}}' =(H_\text{block} -\alpha \mathbb{I})^2$, where the parameter $\alpha$ is adjusted so that the energy levels in the middle turn into the lowest energy levels. This provides a method to calculate all energy levels of a system with more than four neutrinos. 


\subsection{Trotterization}
\label{sec:Trot}

Next, we turn to the calculation of the transition probability amplitude of collective neutrino flavor oscillations. To this end, we need to calculate the real-time evolution of an initial state governed by the unitary time evolution operator $\mathcal{U} =e^{-iH_{\text{flavor}}t}$ with 
\begin{equation}
\begin{split}
    H_{\text{flavor}}=& - \frac{1}{2}\sum_{p=1}^M\omega_p (\cos 2\theta Z_p - \sin 2\theta X_p)\\&+\frac{\mu(r)}{4} ( \bm{X}^2 + \bm{Y}^2 + \bm{Z}^2 )~,\label{eq:FlaHam}
\end{split}
\end{equation}
where we substituted $\Vec{B}=(\sin{2\theta},0,-\cos{2\theta})_{\text{flavor}}$ in Eq.\ \eqref{eq:Ham} for the flavor-basis magnetic field and used $\sin^2(2\theta)=0.1$ for the two-flavor mixing angle which has been obtained from observations of neutrino oscillations~\cite{Cervia2019}. The transition probability amplitude between a given initial state $|\Psi_{\text{initial}}\rangle$ and a final state $|\Psi_{\text{final}}\rangle$ is
\be
P_{fi} (t) = |\mathcal{A}_{fi}|^2 \ , \ \ \mathcal{A}_{fi} (t) = \langle \Psi_{\text{final}}|\mathcal{U} (t) |\Psi_{\text{initial}}\rangle~.
\ee
\begin{figure}[h]
\includegraphics[scale=0.8]{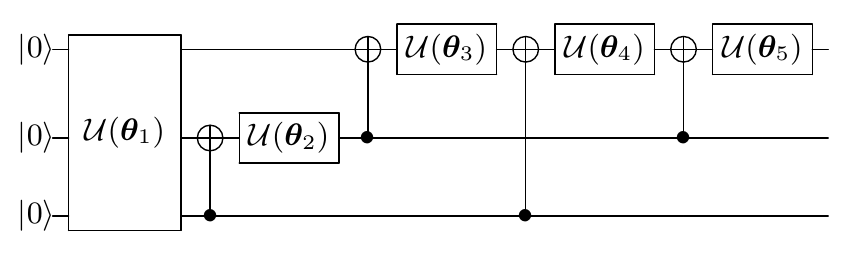}
\caption{The quantum circuit obtained using IBM Qiskit's \textit{isometry} function for the Trotterization of a three-qubit block Hamiltonian of the neutrino system. }
\label{fig:isometryTrot}
\end{figure}
First-order Trotterization is one of the commonly used methods to decompose the unitary time evolution so that the unitary time evolution operator can be approximated by operators which can be expressed in terms of one- and two-qubit operators of the quantum computers. To this end, we split the Hamiltonian into three parts each consisting of commuting matrices, $H = H_X +H_Y + H_Z$, where
\begin{eqnarray}
H_X &=& \frac{\sin 2\theta}{2} \sum_{p=1}^M \omega_p X_p + \frac{\mu (r)}{4} \bm{X}^2 \nonumber\\
H_Y &=&  \frac{\mu (r)}{4} \bm{Y}^2\nonumber\\
H_Z &=& -\frac{\cos 2\theta}{2} \sum_{p=1}^M \omega_p Z_p + \frac{\mu (r)}{4} \bm{Z}^2
\end{eqnarray}
and use
\be
e^{-iH \Delta t}=e^{-iH_X \Delta t}e^{-iH_Y \Delta t} e^{-iH_Z \Delta t}+\mathcal{O}( (\Delta t)^2)
\ee
Then
\be \mathcal{U} (t) \approx \left[ e^{-iH_X \Delta t}e^{-iH_Y \Delta t} e^{-iH_Z \Delta t} \right]^n \ee 
for $t = n\Delta t$. To implement each step on quantum hardware, several gates are needed. Due to the limited decoherence time in NISQ devices, it is often not possible to implement a large number of Trotter step. To circumvent this problem, we used the \emph{isometry} function in the IBM Qiskit library to simplify the circuits. The resulting quantum circuits can be seen in Fig.~\ref{fig:isometryTrot}, where $\mathcal{U}({\bm{\theta}}_i)$ with $i \in \{1,2,3,4\}$ are a series of single-qubit U3($\theta,\lambda,\phi$) gates. The circuits are of this form independently of the number of Trotterization steps. As we add more steps, only the angles change in the circuit of Fig.~\ref{fig:isometryTrot}, which allows us to implement Trotterization on the IBM Q devices. 
\begin{figure*}[ht!]
    \centering
    \includegraphics[scale=0.2]{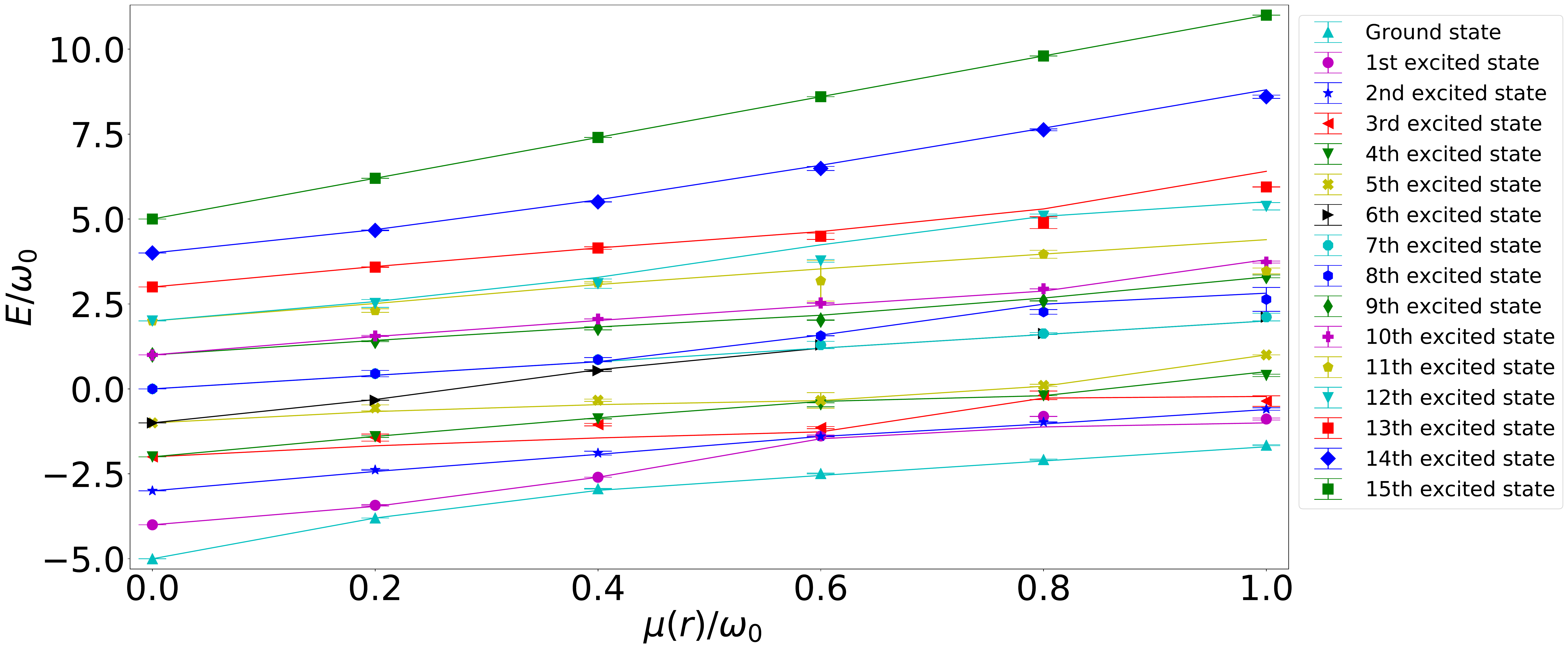}
    \caption{ROEM energy eigenvalues of 4-neutrino system as a function of the radial dependence of the coupling system $\mu(r)/\omega_0$ obtained using QLanczos algorithm. The experiments were run on various IBM Q devices (Athens, Casablanca, Ourense, Rome, Santiago and Vigo) for $N_{\text{runs}}=2$ times and error bars represent $\pm \sigma$. The straight lines represent the exact eigenvalues obtained from exact diagonalization.} 
    \label{fig:energy}
\end{figure*}

Although this simplification is not scalable, it is advantageous for NISQ devices, as the angles change at each Trotter step but the depth of the quantum circuit remains the same. Using this tool is sometimes not straightforward because the quantum circuits obtained using the \textit{isometry} function involve connections between all qubits in the circuit which is not the case for many IBM Q devices. If hardware does not have this kind of connectivity, then addition of extra SWAP gates is required resulting in additional noise in the system. A possible solution for this kind of problem could be using an ion trap quantum computer that has all to all connectivity, thus requiring no additional SWAP gates.

\section{Results}
\label{sec:res}
\begin{figure*}[ht!]
\begin{center}
    \includegraphics[scale=0.6]{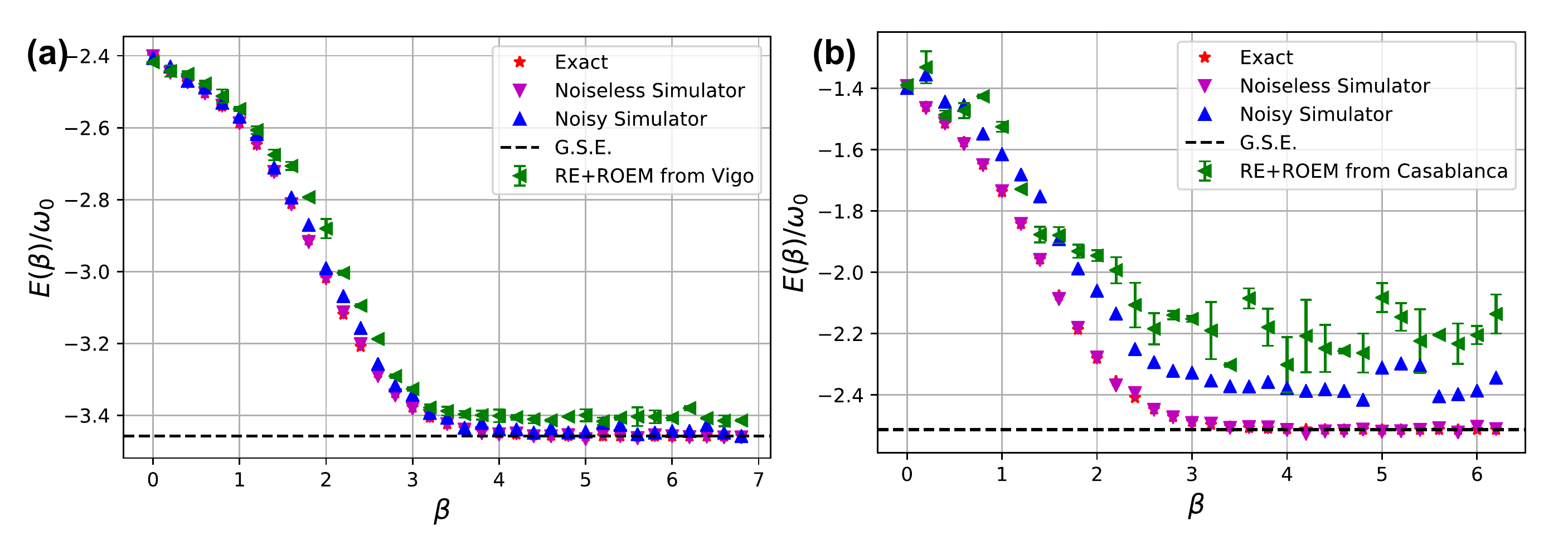}
\end{center}
\caption{\textbf{(a)}: Energy vs. imaginary time calculated exactly using 2-qubit $H_{\text{block}}$ and compared to IBM Q Aer QASM noiseless and noisy simulator with $\mu(r)/\omega_0= 0.2 $ and getting the measured energy expectation values from IBM Q Vigo hardware (data collected on 01/09/2021). \textbf{(b)}: Energy vs. imaginary time calculated exactly using 3-qubit $-H_{\text{block}}$ and compared to IBM Q Aer QASM noiseless and noisy simulator with $\mu(r)/\omega_0= 0.2 $ and using $A[s]$ operators from noisy simulator to get the measured energy expectation values from IBM Q Casablanca hardware (data collected on 01/19/2021). The experiments were run $N_{\text{run}}=2$ times and the error bars represent $\pm \sigma$. The energies converge to the ground state energy level {\bf{(a)}} -3.415 $\pm$ 0.003  with percentage error of 1.3 \% and {\bf{(b)}} -2.205 $\pm$ 0.029 with percentage error of 12.3 \% . } \label{QITEfigs}
\end{figure*}
First, as seen in Fig.\ \ref{fig:energy} we performed calculations with the 4-neutrino system Hamiltonian expressed in the neutrino mass basis to find the eigenvalues as functions of the radial-distance-dependent coupling parameter $\mu(r)/\omega_0$ using the QLanczos algorithm and our experimental data agrees very well with the energy eigenvalues obtained from exact diagonalization. To this end, we divided the Hamiltonian of the system into two- and three-qubit subsystems which helped us obtain results with less hardware noise compared to a four-qubit system needed for the full Hamiltonian. We ran the QITE algorithm as described above for two-qubit circuits on IBM Q Athens, Casablanca, Ourense, Rome, Santiago, and Vigo devices. We ran three-qubit quantum circuits on IBM Q Athens, Casablanca and Lima, switching between devices depending on their availability.

We present two sample data sets obtained with the QITE algorithm in Fig.~\ref{QITEfigs}. Figs.~\ref{QITEfigs}{\bf{(a)}} and \ref{QITEfigs}{\bf{(b)}} depict QITE algorithm results for a two-qubit reduced Hamiltonian block ($H_{\text{block}}$) and three-qubit block with reversed sign ($-H_{\text{block}}$), for $\mu(r)/\omega_0=0.2$, implemented on IBM Q Vigo and Casablanca, respectively. The convergence of the energy expectation value to the ground state energy is evident for the two-qubit $H_{\text{block}}$ in Fig.~\ref{QITEfigs}{\bf{(a)}} starting with initial state $|\Psi_0\rangle=|10 \rangle$.  This convergence is obscured by hardware noise in the 3-qubit case with $-H_{\text{block}}$ starting with initial state $|\Psi_0\rangle=|001 \rangle$, as seen in Fig.\ \ref{QITEfigs}{\bf{(b)}}. We compare the energy expectation value as a function of imaginary time between those obtained from exact diagonalization, noiseless simulator, noisy simulator with quantum hardware noise, and experimental runs performed on the IBM Q Vigo and Casablanca devices. This comparison shows that the noisy simulator with noise parameters from the quantum device is not good enough to reflect all noise present in the quantum hardware. It appears that the cross-talk errors present in quantum hardware are not reflected in the error parameters of the so-called fake back-end noise models. We repeated this procedure with different initial states and values of the coupling $\mu(r)$ with similar results. We then used the output of the QITE algorithm to find all eigenvalues using the QLanczos algorithm.


To obtain Fig.\ \ref{fig:energy}, we performed experiments with $N_{\text{runs}}=2$ runs and $N_{\text{shots}}=8192$ shots per run. The error bars represent $\pm \sigma$. To obtain the experimental energy eigenvalues in Fig.\ \ref{fig:energy}, we used both readout error mitigation (ROEM) and Richardson extrapolation error mitigation techniques. In the implementation of the QLanczos algorithm we discarded eigenvalues if the total uncertainty of two states in Krylov space exceeded a certain value, by demanding $\Delta E \leq \delta$. For three-qubit data, we set $\delta = 1.3$, whereas for 2-qubit data we set $\delta = 0.6$. Then we chose the eigenvalues corresponding to the minimum total uncertainty below this threshold. The threshold was set larger in the three-qubit case because a larger qubit system produces more hardware noise. This resulted in experimental eigenvalues obtained with the aid of the QLanczos algorithm that agreed well with exact eigenvalues obtained from exact diagonalization. 

We performed calculations for various values of the coupling $\mu (r)$. Smaller values yielded better agreement of experimental and exact energy levels. Applying the QLanczos algorithm on $H_{\text{block}}$ and $-H_{\text{block}}$ with a two-dimensional Krylov space yields a couple of lowest and highest eigenvalues in the spectrum, respectively. While this spans the spectrum in the two-qubit case, for a three-qubit system there are a couple of intermediate size eigenvalues which this procedure does not calculate. Fortunately, the remaining eigenvalues correspond to eigenvectors that can be computed exactly using symmetry considerations of the block. We obtain the states $\frac{1}{\sqrt{2}} (|010\rangle-|011\rangle)$ and $\mathcal{N} [(1+ \frac{2}{\mu} )|010\rangle - |000\rangle -2|001\rangle+|101\rangle+2|100\rangle]$, where $\mathcal{N}^{-2} = 11+ \frac{4}{\mu} + \frac{4}{\mu^2}$. We obtained these eigenvalues experimentally by running the quantum circuits for the corresponding eigenstates on IBM Q hardware; see  Fig.~\ref{fig:energy}. 

\begin{figure*}[ht!]
    \centering
    \includegraphics[scale=0.55]{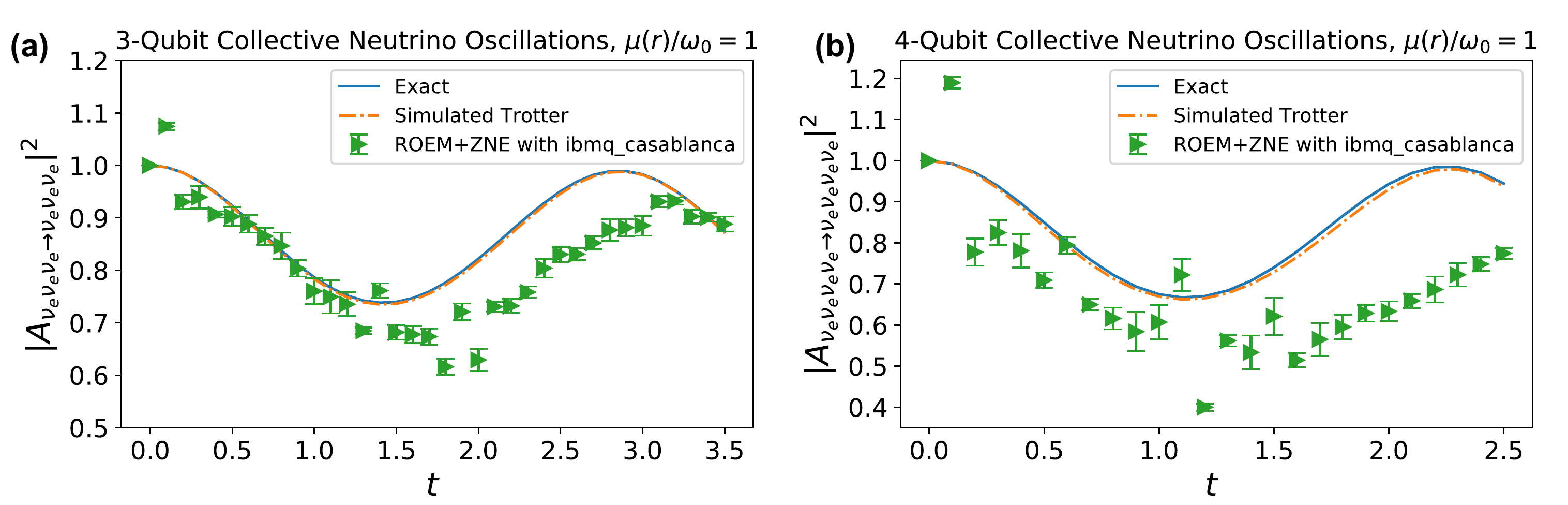}
    \caption{The transition probability amplitude comparison for transition from {\bf{(a)}} $|\text{initial}\rangle=|\nu_e \nu_e\nu_e\rangle $ to $|\text{final}\rangle=|\nu_e\nu_e\nu_e\rangle$, {\bf{(b)}} $|\text{initial}\rangle=|\nu_e \nu_e\nu_e\nu_e\rangle $ to $|\text{final}\rangle=|\nu_e\nu_e\nu_e\nu_e\rangle$ as a function of time between exact time evolution calculations, noiseless simulated Trotterization and values obtained using Trotterization implemented on the seven-qubit IBM Q Casablanca hardware. The error mitigation methods applied are readout error mitigated (ROEM) $+$ zero-noise extrapolation (ZNE) combined. The parameters are chosen such that $\sin^2\left(2\theta\right)=0.10$, $\omega_p=p\omega_0$ and $\mu(r)/\omega_0=1.0 $. The experiments were run {\bf{(a)}} $N_{\text{run}}=5$, {\bf{(b)}} $N_{\text{run}}=6$ times on February, 3-4, 2021 and the error bars represent $\pm \sigma$.} 
    \label{fig:3and4QubitTransProb}
\end{figure*}

We also calculated the transition probability amplitudes for collective neutrino oscillations on a quantum computer. The transition probabilities for 3- and 4-neutrino systems are shown in Figs.~\ref{fig:3and4QubitTransProb} {\bf{(a)}} and {\bf{(b)}}, respectively. We compare exact transition probabilities, transition probabilities from noiseless simulated first order Trotterization, and experimental results obtained from IBM Q hardware. To obtain the time evolution of the initial state we used the Trotterization method discussed in Sec.\,\ref{sec:qc}. The experiments were performed on the IBM Q 7-qubit Casablanca hardware device. We ran the experiments $N_{\text{runs}}=6$ (Fig.~\ref{fig:3and4QubitTransProb}{\bf{(a)}}) and $N_{\text{runs}}=5$ (Fig.~\ref{fig:3and4QubitTransProb}{\bf{(b)}}) times such that each run had 8192 shots. Error bars represent  $\pm \sigma$. Evidently, the main source of error is not algorithmic, but due to quantum hardware noise.

In Fig.~\ref{fig:3and4QubitTransProb}{\bf{(a)}} we were able to obtain the transition amplitudes within the error bars up to $t \approx 1.5$ for 3-qubit case and we can clearly observe the flavor oscillation for this particular case. On the other hand, although most of the exact data points are not within the error bars of the experimental data points, we were able to observe the flavor oscillation even for 4-qubit case (Fig.~\ref{fig:3and4QubitTransProb}{\bf{(b)}}). There are two obvious sources of errors contributing to this situation in addition to other quantum hardware noise. The first one is that
the {\textit{isometry}} function in the Qiskit library provides quantum circuits that require a hardware layout with a connection between all physical qubits. Thus, implementing it on quantum hardware for the transition probabilities, it results in more noise on a four-qubit quantum circuit compared to a three-qubit quantum circuit. In most of IBM quantum devices all-to-all connection between qubits is not available. To trigger interaction between qubits that are not connected, SWAP gates are added to the quantum circuit. Each additional SWAP gate costs 3 CNOT gates on an IBM quantum system increasing the noise in the system significantly. To address this source of error we are going to conduct this experiment on an ion-trapped quantum computer which has all-to-all qubit connectivity and work in this direction is in progress. In addition to the connectivity issue, the quantum circuit is deeper in 4-qubit case than in 3-qubit case and this is another source of error considering the decoherence time in quantum hardware. This source of error can be addressed using different circuit optimization tools such as QSearch~\cite{QSearch:2020} or QGo~\cite{QGo:2021} which might provide a shorter circuit depth and reduce the number of CNOT gates in the quantum circuit. Work in this direction is also in progress. However, the error mitigation strategies we followed allowed us to obtain energy eigenvalues of a 4-neutrino system in very good agreement with the exact eigenvalues. 


For the data presented in Figs.~\ref{fig:energy} and \ref{fig:3and4QubitTransProb} we used both readout error mitigation (ROEM) and zero noise extrapolation (ZNE) to mitigate the error caused by the noise from quantum hardware. To mitigate the error from the measurements, we employed tools included in IBM's \textit{qiskit-ignis} making use of a constrained matrix inversion approach. The response matrix is obtained from the measurements on $2^{N_{\text{qubits}}}$ calibration quantum circuits prepared so that initialization of the qubits was followed by all combinations of the single-qubit $X$ gate. These quantum circuits were bundled with the circuits of interest in the same job. For the error mitigation of the two-qubit gate errors we used the zero-noise extrapolation \cite{Dumitrescu2019} where we added double CNOT gates to the circuit for each CNOT gate in the original quantum circuit in order to increase the noise due to CNOT gate. Then we extrapolated to ``noiseless" expectation values for the energy measurements and count values for the transition probability measurements. We ran the quantum circuits with $r=1,3,5,7,9$, where $r$ is the number of CNOT gates in the quantum circuit for each CNOT gate in the original circuit. Then we fitted the data points by a polynomial of degree $n$ for $n$-qubit quantum circuits ($n=3,4$). For two-qubit circuits, we only used ROEM. 

\section{Conclusion}
\label{sec:con}
In this work, we performed calculations involving many-body collective neutrino-flavor oscillations on a quantum computer using the QLanczos algorithm as well as a first-order Trotterization algorithm. We obtained the energy levels of the Hamiltonian of the system and transition probabilities demonstrating neutrino oscillations on a quantum computer. Working with the collective neutrino oscillations Hamiltonian in the neutrino mass basis, we reduced the Hamiltonian into smaller Hamiltonian blocks to reduce the noise on quantum hardware. With the inclusion of the readout and Richardson extrapolation error mitigation strategies we were able to calculate the energy levels of the system and found them to be in very good agreement with the exact values for a four-neutrino system. We were also able to observe collective neutrino-flavor oscillations in three- and four-neutrino systems on a quantum computer.

Besides the quantum-classical QLanczos algorithm used here, recently a number of novel methods have been proposed for the calculation of the ground state energy as well as higher energy levels of quantum systems. For example, the quantum inverse iteration algorithm~\cite{Kyriienko2020} introduces an approximate ground state which is prepared by a successive application of the inverse Hamiltonian. Another method that was recently proposed is the quantum version of the power method~\cite{Seki2021} in which $\mathcal{H}^n|\psi\rangle$, where $\mathcal{H}$ is the Hamiltonian, is calculated from a time-discretized version of the higher-order derivative of the time-evolution operator $\mathcal{U}(t)=e^{-i\mathcal{H}t}$ using $\mathcal{H}^n=i^nd^n\mathcal{U}(t)/dt^n|_{t=0}$. Also the quantum power method was applied to Krylov-subspace diagonalization. On the other hand, Krylov space methods such as classical Lanczos algorithm has also been proposed as an error mitigation method in~\cite{Suchsland2021}. This method is hardware agnostic and compatible with other error mitigation techniques, and is worth exploring in the context of neutrino oscillations. 

The quantum circuits that were employed in this work generally require connections between all qubits in the circuit. It would be interesting to implement our algorithm on an ion-trapped quantum computer with an all-to-all connectivity and see how our results would improve with less noise. This would also allow us to increase the number of neutrinos in the system. Although all-to-all connectivity might decrease the noise due to addition of extra SWAP gates, this might result in a significant decrease in gate and algorithmic speeds~\cite{Alexeev2021}. It is important to optimize the trade-off between connectivity and speed of the algorithm. An exploration of possible optimization methods is in progress. 
\acknowledgments
We acknowledge useful discussions with A.\ Baha Balantekin. The quantum circuits were drawn using the Q-circuit package \cite{QCircuit}.
This work was supported by the Quantum Information Science Enabled Discovery (QuantISED) for High Energy Physics program at ORNL under FWP number ERKAP61 and used resources of Oak Ridge Leadership Computing Facility located at ORNL, which is supported by the Office of Science of the Department of Energy under contract No.\ DE-AC05-00OR22725.
 The authors acknowledge use of the IBM Q for this work. The views expressed are those of the authors and do not reflect the official policy or position of IBM or the IBM Q team. G.S.\ is supported by ARO grant W911-NF-19-1-0397 and NSF grant OMA-1937008.




\twocolumngrid

 \onecolumngrid

\twocolumngrid

%
\begin{appendix}
\section{QITE Algorithm}
\label{sec:QITE}
In this appendix we briefly discuss the quantum imaginary-time evolution (QITE) algorithm which provides the basis for the quantum Lanczos (QLanczos) algorithm that is used to calculate the energy spectrum of the collective neutrino system that we studied.

In the QITE algorithm, the non-unitary imaginary-time evolution operator $\mathfrak{U} =e^{- \beta H_\text{block}}$ for imaginary-time evolution of an initial state can be expressed as 
\be
|\Psi(\beta)\rangle=c_n (e^{-\Delta \tau H_\text{block}})^n|\Psi_0\rangle
\ee
after dividing the imaginary-time evolution into $n$ small steps with $\Delta \tau=\frac{\beta}{n}$. We included a normalization constant $c_n=\frac{1}{\sqrt{\langle \Psi_0|\mathfrak{U}^2|\Psi_0\rangle}}$. The $s$th step of this imaginary-time evolution is approximated by real-time evolution in terms of a unitary operator as
\be
|\Psi_s\rangle=\frac{c_s}{c_{s-1}}e^{-\Delta \tau H_\text{block}}|\Psi_{s-1}\rangle \approx e^{-i\Delta \tau A[s]}|\Psi_{s-1}\rangle
\ee
where $c_s$ is a normalization constant that can be calculated recursively starting with $c_0=1$. The unitary update operators can be written as 
\be
A[s]=\sum_{i_0,\dots,i_{n_q-1}}a_I[s] \sigma_{I} ~.\label{eq:As}
\ee
where $I=\{i_0,\dots, i_{n_q-1}\}$, $n_q$ is the number of qubits for the reduced block Hamiltonian, and $\sigma \in \{X,Y,Z\}$ are Pauli operators.  The coefficients $a_I[s]$ are calculated by solving the linear system of equations $({\bm{\mathcal{S}}}+{\bm{\mathcal{S}}}^T)\cdot {\bm{a}}={\bm{b}}$ up to order $\mathcal{O}(\Delta \tau^2)$, where
\be
\mathcal{S}_{I}=\langle \sigma_{I} \sigma_{I'} \rangle \ , \ \
b_{I}=-i\sqrt{\frac{c_{s-1}}{c_s}}\langle \sigma_{I} H_\text{block} \rangle \label{eq:bmat}
\ee
These expectation values are calculated with respect to the state in the previous QITE step, $|\Psi_{s-1}\rangle$. The exponentially growing size of the matrices ${\bm{b}}$ and ${\bm{\mathcal{S}}}$ challenges the scalability of the algorithm. In our case, certain properties of the Hamiltonian, such as real matrix elements, help us reduce the size of these matrices. Due to the large number of measurements required for the calculation of matrix elements on quantum hardware accessible through cloud services, for a timely completion of calculations some of our measurements were performed using a noisy simulator instead. This resulted in a reduction of overall error but did not affect results significantly, as was ascertained by a comparison of results between sample quantum hardware and simulated runs.

We point out that circuit bundling, which is sometimes used to minimize the size of jobs submitted for quantum hardware runs, is not an option in our case, because the outcome of each step is used in the following step. Moreover, the QITE algorithm requires single- and two-qubit gates at each step resulting in a quantum circuit depth exceeding the decoherence time in NISQ devices. We took steps to shorten the quantum circuits by adapting methods we introduced in our earlier work~\cite{YeterAydeniz2020, YeterAydeniz2021Chem}.

In detail, we started by determining the number of QITE steps in the implementation of our algorithm. We selected the number of steps that yielded convergence of energy expectation values within $\epsilon=0.001$ of the ground state energy. We used this criterion to estimate the required number of steps numerically. Then we ran experiments on a noisy simulator of the quantum device that produced the matrix elements for $\bm{S}$ and $\bm{b}$ at each step, and used them to solve classically the system of equations $({\bm{\mathcal{S}}}+{\bm{\mathcal{S}}}^T)\cdot {\bm{a}}={\bm{b}}$ obeyed by the coefficients $a_I[s]$ introduced in Eq.\ \eqref{eq:As}.
Next, to shorten the depth of the quantum circuit so it is manageable by NISQ devices, we employed the \textit{isometry} function from the IBM Qiskit library which uses the algorithm developed in~\cite{Iten2020} where a given isometry is decomposed into single-qubit and Controlled-NOT (CNOT) gates with the aim of having the least number of CNOT gates. We used this \textit{isometry} function to simplify the circuits of the state at the $s$th step,
\be
    |\Psi_s\rangle=e^{-i\Delta \tau A[s]}e^{-i\Delta \tau A[s-1]}\cdot \cdot \cdot e^{-i\Delta \tau A[1]}|\Psi_0\rangle~.
\ee
It turns out that the resulting quantum circuit at each QITE step has the same gate components but with different rotation angles making the implementation of the QITE algorithm possible on NISQ devices. We ran these quantum circuits on IBM Q quantum hardware to obtain experimental energy expectation values. 
\end{appendix}
\end{document}